# High sensivity and full-circle optical rotary sensor for non-cooperatively tracing wrist tremor with nanoradian resolution

Xin Xu, Zongren Dai, Yifan Wang, Mingfang Li, and Yidong Tan*

*State Key Laboratory of Precision Measurement Technology and Instruments, Department of Precision Instruments, Tsinghua University, Beijing 100084, China*
*Corresponding author: tanyd@tsinghua.edu.cn*

**An optical rotary sensor based on laser self-mixing interferometry is proposed, which enables noncontact and full-circle rotation measurement of non-cooperative targets with high resolution and sensitivity. The prototype demonstrates that the resolution is 0.1μrad and the linearity is 2.33×10$^{-4}$. Stability of the prototype is 2μrad over 3600s and the repeatability error is below 0.84°under 9-gruop full-circle tests. The theoretical resolution reaches up to 16nrad. Random rotation has been successfully traced with a bionic hand to simulate the tremor process. Error analysis and limitation discussion have been also carried out in the paper. Although the accuracy needs further improvement compared with the best rotary sensor, this method has its unique advantages of non-cooperative target sensing, high sensitivity and electromagnetic immunity. Hence, the optical rotary sensor provides a promising alternative in precise rotation measurement, tremor tracing and nano-motion monitoring.**

## 1. INTRODUCTION

Rotation measurement plays a crucial role in industrial applications and frontier scientific research, such as automatic control, robotics, precision machining, mask aligner and gravitational wave detection [1-5]. Besides the high stability, repeatability, applicability and low complexity, which are important factors for the rotary sensors, the non-contact measurement, full-circle sensing range, high resolution and accuracy are required in the practical applications. Many rotary sensors have been developed in the last several decades based on various working principles, such as the mechanical, electronic and optical methods. Currently, mechanical and electronic sensors are mostly utilized in the rotation measurement. Commercial mechanical turntables and sensors are usually limited by the resolution and accuracy. Electromagnetic and capacitive rotary encoders are able to provide accurate angle measurement even in harsh environment [6,7]. And the resolution can reach micro-radian level and the measuring error is lower than ±5arcsecond over the full-circle range after artificial correction [8,9]. However, the biggest drawback of the electronic rotary sensors is that the sensing head needs to be fixed on the rotary target. This limitation of non-cooperative target measurement severely narrows the application range of electronic rotary sensors.

As to optical metrology, it has attracted increasing attention due to its nondestructive detection, tracking to the wavelength source and strong electromagnetic immunity. Normal interferometric methods need two beams or more to decouple the motions in multi-degree-of-freedom for the angle measurement [10,11]. Some other optical methods can also measure the angle variation , including auto-collimatic method [12], total internal reflection method [13], grating-based interferometer [14], birefringence heterodyne interferometry [15], surface plasmon resonance [16], self-mixing interferometry [17], and dual-comb interferometry [18]. These methods reach a high resolution, for instance, 0.001arcsecond angular displacement has been measured experimentally [10], but the sensing range of the above interferometric methods is strictly limited within several degrees. Circular optical grating encoder is the most precise method of full-circle rotation measurement at present, in which the high precision is owing to the modern grating manufacture [19]. It means that the rotary sensors based on optical grating are expensive and dedicate. More importantly, most optical rotary sensors fail to non-cooperatively measure the rotation without contacting or mirror-aided. The sensing part must attach to the rotary target or the assisted mirrors are used to acquire rotary signals during the measurement. Only a few ways can achieve noncontact or no-mirror-aid measurement, such as Doppler velocity [20] and digital cameras [21]. Nevertheless, their precision and range are not acceptable. Therefore, achieving the non-cooperative rotation measurement with high precision, high resolution, full-circle, non-contact and easy-operation at the same time is demanding and meaningful.

Recently, laser self-mixing interferometry(LSI) has been successfully applied in many fields, including displacement/velocity measurement, laser eavesdropping and biological imaging [22,23]. Its high sensitivity and auto-collimation could provide a noncontact measurement of non-cooperative targets [23], which matches well with the demands in rotating measurements. Consequently, in this paper, we designed an optical rotary sensor(ORS). The pioneering method is able to measure the full-circle rotation with high resolution and stability, which minimizes the disruption of installation or added mirrors. Also, high sensitivity of laser self-mixing interferometry promises wide applicability for rotating stages of different materials. Furthermore, we recovered the wrist tremor through the proposed ORS by measuring the random rotation of a bionic hand. Hand/wrist tremor is the basic feature of Parkinsonism, as it is of great significance to trace it in the early stage. Owing to the advantages of noncontact measurement, low power and high sensitivity, the proposed ORS has promising applicability to trace the tremor. In the paper, the design and principle of the ORS is introduced in the Part 2. The proposed prototype is tested with the high-precision rotation instruments for the stability, resolution, linearity, repeatability and tremor trace experiments, as shown in Part 3. Error analysis and conclusions have been brought out in the final part.

## 2. DESIGN AND BASIC PRINCIPLE

### A. ORS Structure

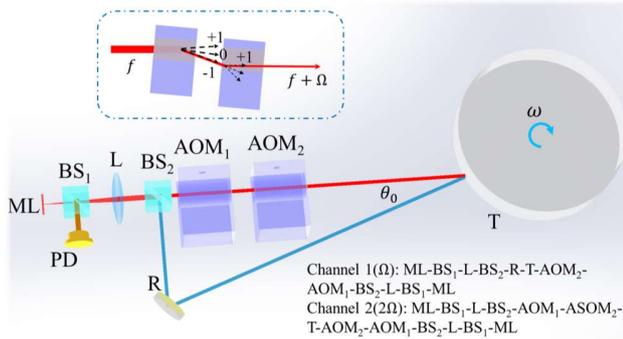

Fig. 1. Schematic diagram of the optical rotary sensor.
ML: microchip laser; BS: beam splitter; L: lens; R: reflector;
AOMs: acoustic-optic modulators T: target; PD: photodetector;

The optical rotary sensor structure is designed as shown in Fig. 1. Paths of dual-channel sensing beams have been labeled. A pair of AOMs provide a heterodyne frequency-shifted modulation of $\Omega$. Sensing beams are incident onto the rotary target, and the reflected or scattered part will feedback into the inner cavity. The feedback signals will cause self-mixing effect on the laser output power, which can be expressed as [24]

$$\frac{\Delta I(\Omega)}{I} = \kappa_1 G(\Omega)\cos(2\pi\Omega t - \phi_{01} + \Delta\phi_1)$$
$$\frac{\Delta I(2\Omega)}{I} = \kappa_2 G(2\Omega)\cos(4\pi\Omega t - \phi_{02} + \Delta\phi_2)$$

(1)

where $\kappa$ is the effective reflection coefficient of the external target, $\phi_{01}$ and $\phi_{02}$ are the initial phase of $\Omega$ and $2\Omega$ channel, respectively. $G$ is the gain function. $\Delta\phi_1$ and $\Delta\phi_2$ represent the external phase of the two different channels.

In microchip lasers, the modulation gain $G$ can reach up to $10^6$, which means that the reflectivity of the measured target is as low as $10^{-12}$ when a relative amplitude of a 100% modulation is achieved [23,24]. Therefore, it makes the proposed ORS applicable for the rotation measurement of the non-cooperative targets without the aimed-mirrors

### B. Sensing principle

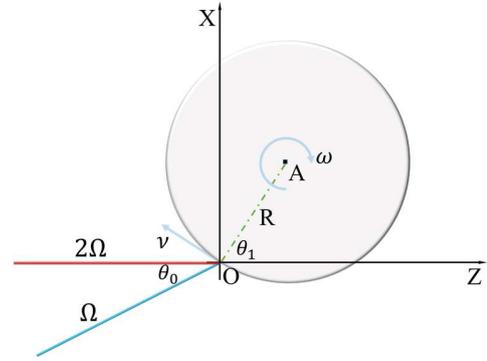

Fig. 2. Rotation sensing principle. O represents the coordinate origin.

Based on the frequency-shifted self-mixing modulation and specific beam-separation structure, we propose a rotation measurement method for column targets. Taking the incident point as the origin O and the channel $2\Omega$ beam propagation direction as the positive z-axis, a plane rectangular coordinate system is established in Fig. 2. $v$ represents the rotary velocity of the target. $\theta_0$ devotes the angle between two sensing beams and $\theta_1$ is the angle between OA and the z-axis, in which the point A is the center of the rotation target. It is worth to note that we consume the two sensing beams and the rotary plane of the target is coplanar. The non-coplanar condition will be discussed in the error analysis part.

Clearly, the rotation of the target will lead to the frequency shift for the two sensing beams, namely the Doppler effect. Thereby it will cause the phase variation $\Delta\phi_1$ and $\Delta\phi_2$ respectively:

$$\Delta\phi_1 = 2\pi\int \Delta f_1 dt = \frac{2\pi f_0}{c}\int(\sin(\theta_1)+\sin(\theta_1-\theta_0))vdt$$
$$\Delta\phi_2 = 2\pi\int \Delta f_2 dt = \frac{2\pi f_0}{c}\int 2\sin(\theta_1)vdt$$

(2)

where $f_0$ is the laser frequency and c is the light velocity in the vacuum.

The rotation can be deduced without the incident angle $\theta_1$ as

$$\theta = \frac{\int vdt}{R} = \frac{\lambda}{2\pi} \frac{\sqrt{\left[\frac{2\Delta\phi_1 - (1+\cos(\theta_0))\Delta\phi_2}{2\sin(\theta_0)}\right]^2 + \left[\frac{\Delta\phi_2}{2}\right]^2}}{R} \quad (3)$$

$\theta_0$ can be calibrated in the experiment as the foreknown parameter and $\lambda$ is the laser wavelength [25]. Thereby, the rotation is linked to the changing phases of two incident beams. The proposed method can achieve the noncontact full-circle rotation measurement without assistant mirrors based on its special sensing principle. This feature distinguishes the proposed ORS from other optical methods.

## 3. EXPERIMENTAL SETUP AND RESULTS

### A. Prototype

There are three main parts of the optical rotary sensor, including optical sensing head, signal processing and host computer.

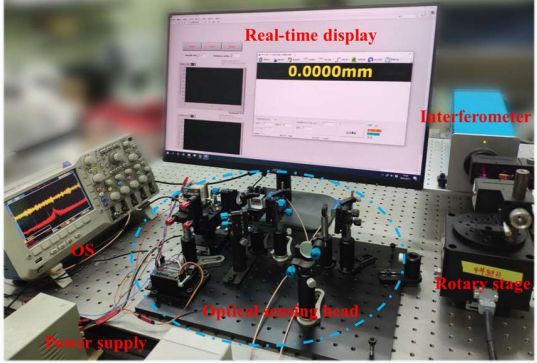

Fig. 3. The optical rotary sensor prototype.

A solid-state microchip Nd:YVO₄ laser is employed for its high gain under frequency-shifted feedback modulation. The laser output beam is linearly polarized and 10.4mW of power in a TEM00 mode. The wavelength stability is $2\times10^{-8}$/h and the power stability is 0.1%/2h [26]. Owing to the extremely short photon lifetime compared to the fluorescence lifetime of Nd:YVO₄ crystal, the optical rotary sensor using a solid-state microchip laser has a ultrahigh detecting sensitivity (or signal to noise radio, SNR) [27]. Therefore, the proposed ORS can measure different targets with low reflectivity. This feature will be demonstrated in detail in the discussion part.

The laser output is separated by a beam splitter (BS$_1$, 0.96:0.04). A lens (L, f=50mm) is placed after BS$_1$ for beam collimation. The key designs of the ORS lie in the beam separation and heterodyne modulation, as introduced in the Part 2. The reflective beam of BS$_1$ is detected by a silicon-based photodetector, which contains the modulated signals and the relaxation oscillation [22]. The whole optical sensing head is 300*200*200mm, which is smaller than many other optical systems of rotation sensing. It is noted that all the employed optical surfaces have been coated with 1064nm anti-reflection films to decrease the effects of parasitic feedback. In the experimental setup, the testing target is an anodized aluminum cylinder, the rotary radius R is 50mm and the angle $\theta_0$ is calibrated as 47.98° [25].

Fig. 4. shows the flow of signal processing and sensing display. The obtained signals and the reference signals from the AOM drivers are directly sent to a lock-in amplifier (Zurich Instrument, HF2LI). Phase will be acquired into the host computer by a data acquisition board (NI, USB-6211). Therefore, a real-time rotary sensor is developed containing the angular displacement and velocity with a LabView block diagram.

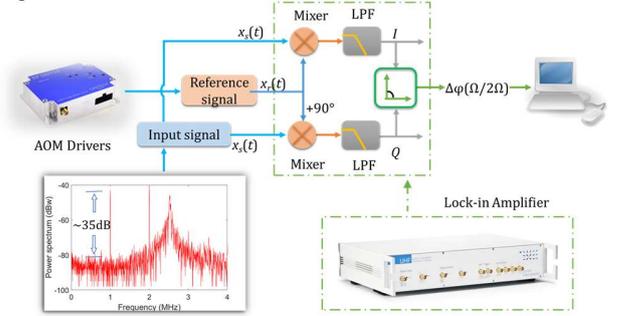

Fig. 4. Dual-channel phase demodulation of the optical rotary sensor. Mixer: frequency mixer; LPF: low-pass filter.

### B. Experimental results

*1. Stability*

To test the rotation measurement performance, the stability results of the prototype have been firstly obtained. Fig. 5. presents the angular displacement when the rotor keeps stationary. One should note that the results are measured after 1-hour power-on preheating. We first take out a short-term stability test in comparison to the interferometer (LeiCe Technology, LH-3000, resolution 0.01arcsecond, 100arcsecond accuracy over ±10°). As Fig. 5a shows, the rotation drift is below 80nrad/min. The theoretical resolution can be calculated from the standard deviation of the stability data [14], which is 16nrad. In this case, nanoradian resolution and ultra-high stability of the rotation sensing can be achieved.

As to long-time stability, we conduct 3600s stationary angular displacement measurement in four days. The Amplitude Spectrum Density(ASD) results are also given, as illustrated in Fig. 5c. The results demonstrate that the ORS has a distinguished stability during the short&long-term test. It is remarkable that the spectrum density of rotation can reach up to 10nrad/Hz$^{1/2}$@1Hz, which is qualified in some precise rotation measurement applications [10].

*2. Rotation resolution*

To determine the resolution of the proposed sensor, step/square-wave tests through a high-precision rotary stage (PI P-562.6CD, 0.1μrad resolution) have been performed. The corresponding results are shown in Fig. 6. Using the PI capacitive sensor as reference, the proposed method proves to be able to distinguish 0.1μrad(0.02arcsecond) angular displacement. It is noted that the short-term drift is unavailable but factual, which is caused by and environmental perturbation, including air convection and temperature varying.

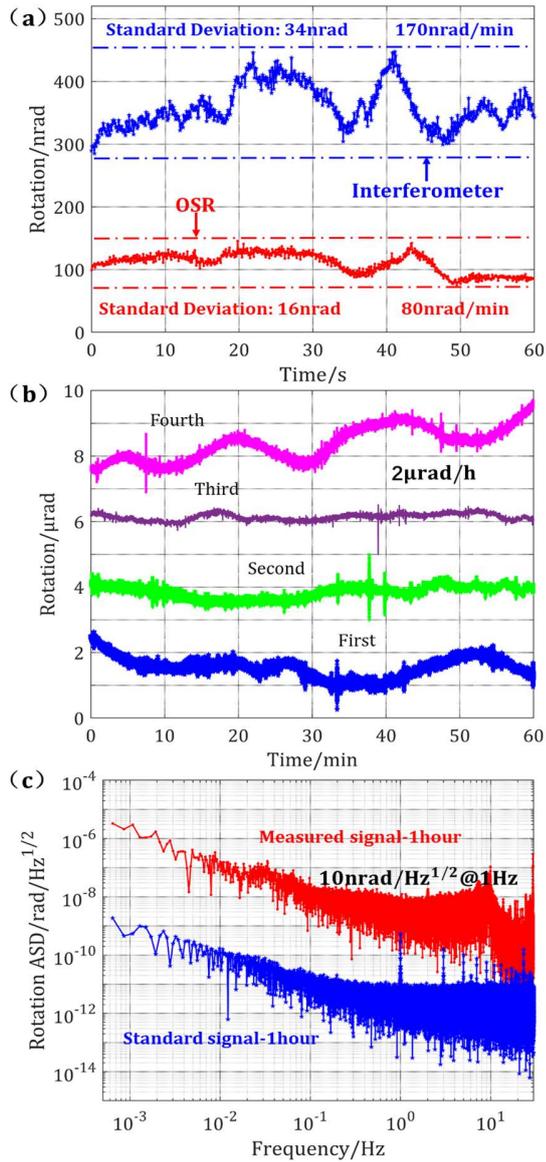

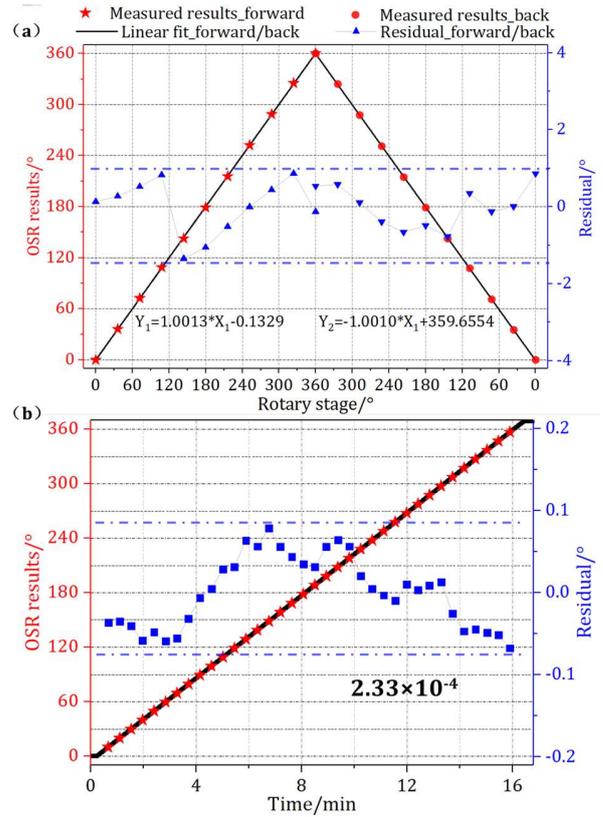

Fig. 5. Stability tests. a. short term (1 min) b. long term (1 hour) c. rotation amplitude spectrum density, optical signal from the actual target & standard signal from the AOMs driver. Data in a/b have been shifted up for convenient display, or the units of the longitudinal axis in a/b are both 100nrad/div.

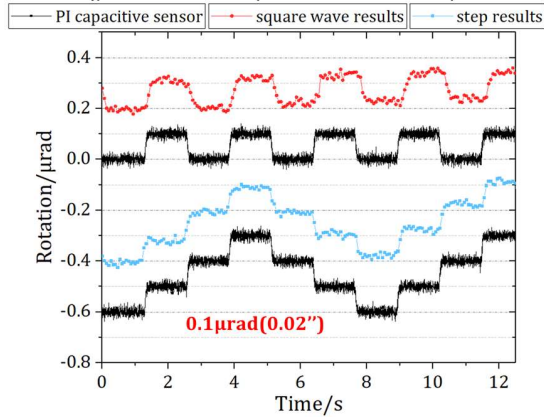

Fig. 6. Resolution tests of square wave and step rotation.

Fig. 7. Linearity test. a. 36° step of one circle test in the forward and back direction b. a continuous rotation beyond 360°. Stars represent the fitting points of linearity.

## 3. Linearity and measuring range

Linearity is another important indicator for the rotary sensors. To better test the real linearity, we conduct a 36° step-by-step rotation of a full-circle and a continuous rotation motion beyond 360° with a commercial rotary stage (Beijing Optical Century Instrument, RS211, 10arcsecond accuracy over 360° and 0.33arcsecond resolution). The nonlinear error in the 36°-step rotation test is assumed to be caused by the start/stop of the rotary stage. The obtained results are presented in Fig. 7, showing that the proposed sensor has a linearity of $2.23\times10^{-4}$ over 360° range.

## 4. Repeatability test

Repeatability is defined as the standard deviation of the measured position error between the starting position and the final position [15]. Definitely, a smaller deviation indicates that the rotary sensor has a higher repeatability. To verity the repeatability of the proposed ORS, the rotation stage is operated 9 times forward and backward with a 36° step during the 360° range. As can been seen in Fig. 8, the experimental results obtained by the proposed sensor and the rotation stage are in good agreement with each other. The standard deviations of 9 groups are lower than 0.84°. The test results clearly demonstrate that the proposed ORS possesses a high measuring repeatability for the rotation sensing.

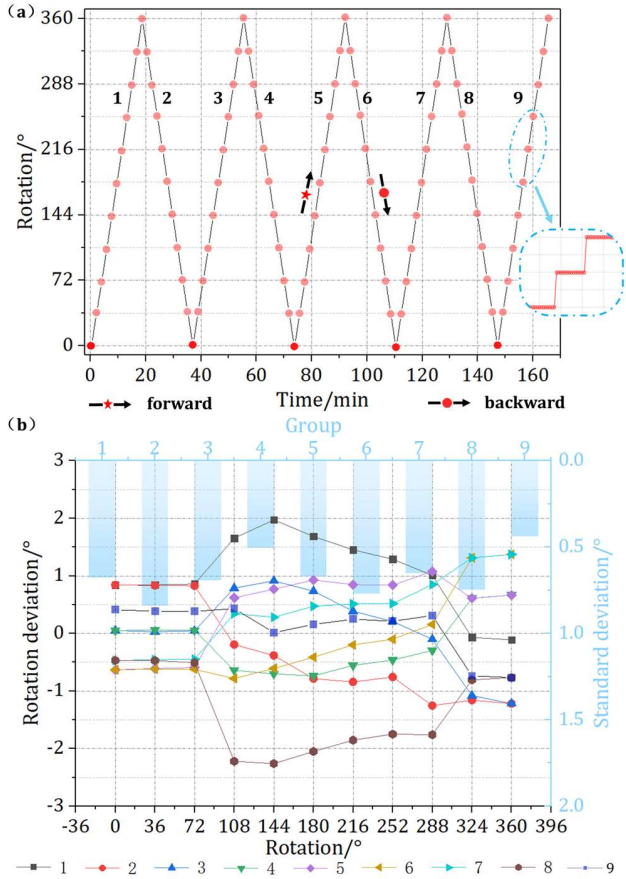

Fig. 8. Repeatability test. a. 9 times test with 36° step of full-circle rotation. b. error results and standard deviation of the repeatability test. The dotted line in Fig. 8b represents the difference between measured results and the 9-group average value, while the column represents the error's standard deviation.

*5. Tremor tracing*

Hand/wrist tremor may indicate Parkinsonism in the early stage. The ORS can be employed to trace wrist tremor and recover the rotation information during the tremor process. In the experiment, the rotary stage produces a random rotation motion of a bionic hand, as shown in Fig. 9. Compared with the reference results from the capacitive sensor in the PI stage, the ORS can measure the random rotation at the micro-motion level. In this case, the proposed ORS exhibits an excellent traceability for the simulated tremor of the robotic hand, which may provide a precautionary diagnosis method for some tremor diseases.

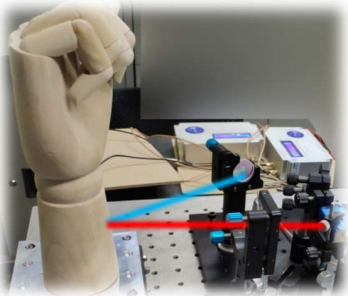

Fig. 9. Experimental setup of tracing the wrist tremor.

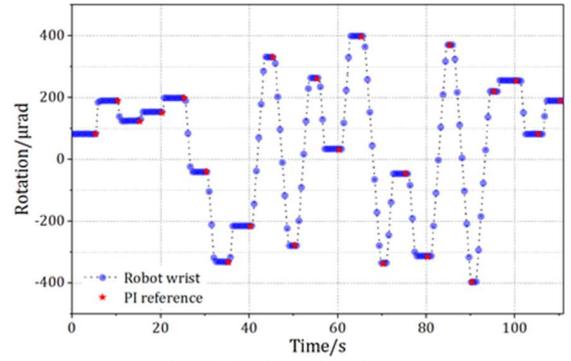

Fig. 10. Recovered results of simulated tremor motions. Five-pointed star gives the reference results from the capacitive sensor in the PI stage.

## 4. DISCUSSION AND CONCLUSION

### A. Error analysis

In the prototype, the sensing beams are assumed as parallel to the rotation plane. However, there is a misalignment (incorrect pitch angle) between the rotation plane and two sensing beams. Thus, the final calculation should be modified as follows:

$$\theta = F(\lambda, R, \beta, \Delta\phi_1, \Delta\phi_2, \theta_0)$$
$$= \frac{\lambda}{2\pi} \times \frac{\sqrt{\left[\frac{2\Delta\phi_1 - \Delta\phi_2(1+\cos(\theta_0))}{2\sin(\theta_0)}\right]^2 + \left[\frac{\Delta\phi_2}{2}\right]^2}}{R \cdot \cos(\beta)} \quad (4)$$

where $\beta$ devotes the pitch angle.

Based on the designed prototype, we bring out the uncertainty analysis of the rotation measurement. More details can be found in the Supplemental document. The results in Fig. 11 demonstrate that the errors mainly come from the uncertainty of $R$ and $\theta_0$. The final relative error is $2.29\times10^{-4}$ over 360° when the confidence factor is 2 through the numeral calculation. It is in good agreement with the experimental results of the linearity. This analysis helps to figure out the main error sources. Accordingly, the proposed ORS is believed to be limited by the uncertainty of $R$ and $\theta_0$, which needs further improvement in the future research.

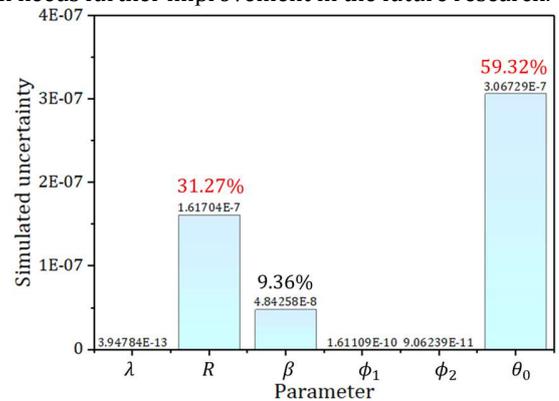

Fig. 11. Relative error results of the proposed ORS.

### B. Discussion

*1. Testing targets made of different materials*

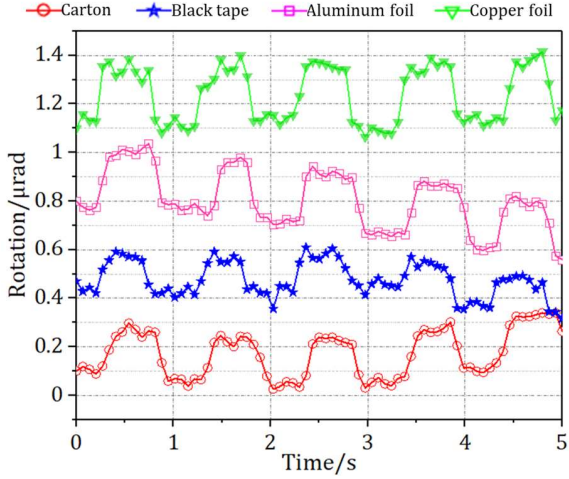

Fig. 12. Resolution tests of different targets.

To further verity the dynamic performance of the proposed sensor, rotation tests with different targets of various materials have been launched out. The obtained results are presented in Fig. 12. It can be seen that the sensor can still distinguish rotation even when the target is of low reflectivity, which verifies the high sensitivity of the proposed ORS based on laser self-mixing interferometry. In addition, all the materials used for tests exhibits a 0.2μrad resolution of angular displacement. Therefore, the proposed ORS has a wide detecting capability and is promising to be used in many applications.

*2. Limitation of the measurement velocity*

Doppler shift caused by the rotation can not exceed the signal processing bandwidth, thus imposing a limitation on the rotary velocity range. According to the Equation 2, the Doppler shift of the two sensing beams can be deduced as:

$$\Delta f_{D1} = (\sin(\theta_1) + \sin(\theta_1 - \theta_0))\frac{v}{\lambda}$$
$$\Delta f_{D2} = 2\sin(\theta_1)\frac{v}{\lambda} \quad (5)$$

When the rotary radium is constant, the highest angular velocity is determined by the sampling bandwidth and the incident angle $\theta_1$. To obtain a higher rotary velocity and a better SNR, the incident angle $\theta_1$ is set zero in the experiment. The experimental bandwidth is set to be 2kHz, as broader bandwidth means more low frequency noise. In this case, the rotation velocity is allowed to be measured below 1.64°/s. The measurable angular velocity range can be improved by increasing the demodulation bandwidth to megahertz level with a phasemeter (PT-1313B).

$$\omega_{max} = \frac{v_{max}}{R} = \frac{1}{R}\min\left\{\left|\frac{\Delta f_D \lambda}{\sin(\theta_1) + \sin(\theta_1 - \theta_0)}\right|, \left|\frac{\Delta f_D \lambda}{2\sin(\theta_1)}\right|\right\} \approx 1.64°/s \quad (6)$$

*3. Summary of the proposed ORS and comparison*

To better conclude this method, a performance summary of the proposed ORS is given in Table. 1. The comparison with other methods of rotation measurement are also presented in Fig.13. In general, the proposed ORS proves that the advantage over electronic sensors lies in the high resolution of $10^{-8}$ rad, and over other optical sensors lies in the full-circle range and lower relative error. Notably, the ORS enables non-cooperative rotary measurement due to the high sensitivity of laser self-mixing interferometry. This characteristic is promising for many applications, such as satellite posture control, robot gesture recognition and hand tremor monitoring.

Table. 1 summary of the proposed ORS

| Optical rotary sensor | Value |
|---|---|
| Target | Non-cooperative |
| Power | 10.4mW |
| Stability | 80nrad@1min, 2μrad @1h |
| Resolution | 0.1μrad (0.02") @experiment<br>16nrad @theory |
| Linearity | 2.33×10$^{-4}$@360° |
| Repeatability | 0.84°@360° (9 times) |
| Angular displacement range | 360° |
| Angular velocity range | 0-1.64°/s (R=50mm) |

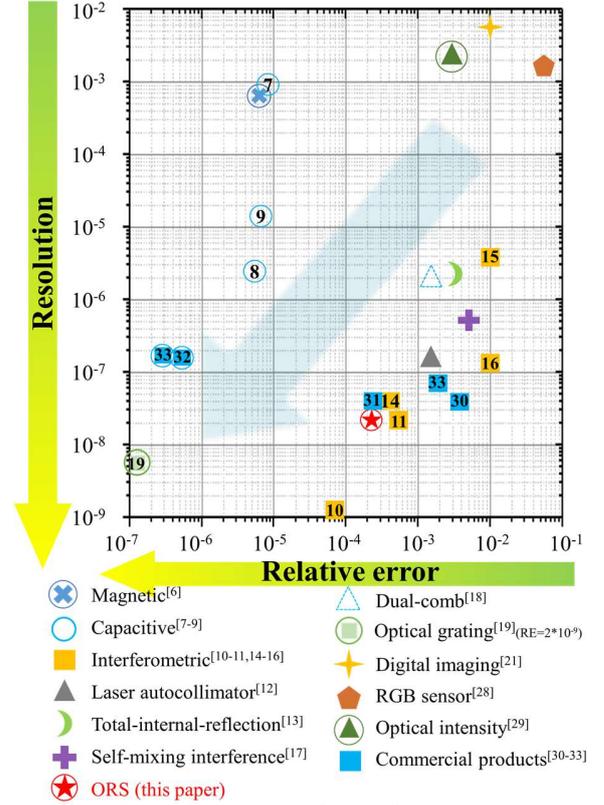

Fig. 13. Performance comparison of different rotary sensors.

**C. Conclusion**

An optical rotary sensor of non-cooperative targets with high sensitivity, full-circle range and nanoradian resolution has been proposed and tested. Experimental results demonstrate that 360° rotation sensing with a measured 0.1μrad resolution and 2.33×10$^{-4}$ precision has been achieved in the laboratory condition. The theoretical resolution can be as low as 16nrad calculated from the stability data. Preliminary attempt has been made to trace the hand/wrist tremor with the proposed ORS. The main error sources during the rotation sensing process are assumed to be from the uncertainty of the rotary radius and the angle of two sensing beams. We also discussed the various sensing performances of targets with different materials and the sensing speed limitation of the present prototype. At last, a summary and comparison have been given, indicating that this method possesses unique advantages and thus promising application potential. Although having a lot of work to optimize, we will focus on the ORS promotion of sensing precision and rotary speed limitation in the future research, hoping to breakthrough in the field of robot gesture sensing, tremor tracing and nano-motion measurement.

**Funding.** National Key Research and Development Program of China (2020YFC2200204); National Science Fund for Excellent Young Scholars (51722506).

**Acknowledgments.** Xin Xu thanks Professor Zeng Lijiang, Li Yan and Zhou Bin from Tsinghua University for the help providing the high-precision rotary instruments.

**Disclosures.** The authors declare no conflicts of interest.

**Supplemental document.** See Supplement 1 for supporting content, including the deduction process of rotation and the uncertainty analysis.